\documentclass[12pt]{iopart}
      \newcommand{\beq}{\begin{equation}}
      \newcommand{\eeq}{\end{equation}}
      \newcommand{\beqa}{\begin{eqnarray}}
      \newcommand{\eeqa}{\end{eqnarray}}
      
      \newcommand{\nn}{\nonumber}

      \newcommand{\bra}{\left\langle}
      \newcommand{\ket}{\right\rangle}
      
      \newcommand{\del}{\partial}

      \newcommand{\al}{\alpha}
      \newcommand{\be}{\beta}
      
      \newcommand{\de}{\delta}
      \newcommand{\ep}{\epsilon}

      \newcommand{\la}{\lambda}
      
      \newcommand{\La}{\Lambda}

\renewcommand{\(}{\left(}
\renewcommand{\)}{\right)}
\usepackage{graphicx}
\begin{document}
% Journal identifier can be put here if required, e.g.
%\jl{14}

\title{Stability of a fixed point in the replica action for the
random field Ising model}

\author{Hisamitsu Mukaida\dag \footnote[3]{mukaida@saitama-med.ac.jp}
and
Yoshinori Sakamoto\ddag \footnote[4]{yossi@phys.cst.nihon-u.ac.jp}}

\address{\dag\ Department of Physics, Saitama Medical College,
981 Kawakado, Iruma-gun, Saitama, 350-0496, Japan}

\address{\ddag\ Department of Physics, College of Science and Technology,
Nihon University, 1-8-14, Kanda-Surugadai, Chiyoda-ku, Tokyo, 101-8308,
Japan}

\begin{abstract}
We reconsider stability of the non-trivial fixed point 
in $6-\ep$ dimensional effective action for the random 
field Ising model derived by Br\'{e}zin and De Dominicis.
After expansion parameters of 
physical observables are clarified,  
we find that the non-trivial fixed point in $6-\epsilon$
dimensions is stable, contrary to the argument by 
Br\'{e}zin and De Dominicis.   
We also computed the exponents $\nu$ and $\eta$  by the 
$\ep$ expansion.  The results are consistent with the 
argument of the dimensional reduction at least in the 
leading order.

%We reexamine the effective action for the
% random field Ising model derived by Br\'{e}zin and De Dominicis.
%We discuss how physical observables are expanded
%by  strength of the random field $\Delta$
%and the $\phi^4$ coupling constants.
%Moreover, we reconsider the instability
%of the non-trivial fixed point in $6-\epsilon$ dimensions. 
%It is found that the non-trivial fixed point in $6-\epsilon$
%dimension is stable, contary to the argument by 
%Br\'{e}zin and De Dominicis.   
%it is consistent with
%the argument of the dimensional reduction through the computation of
%the leading order of the $\epsilon$-expansion for the critical exponents
%$\nu$ and $\eta$ at the non-trivial fixed point.

%We find a non-Gaussian fixed point where the $\phi^4$ couplings
%in the action have various scaling  dimensions.   The  correlation-length
% exponent in $d=6-\ep$ has the value consistent with the argument
% of the dimensional  reduction  in the leading order.
\end{abstract}

\pacs{05.50$+$q; 64.60Fr}

% Uncomment for Submitted to journal title message
%\submitted

% Comment out if separate title page not required
%\maketitle

\section{Introduction}
Recently, Br\'ezin and De Dominicis derived effective scalar field theory
for the Ising model in a Gaussian random field within the
replica formalism \cite{bd}.  They showed that
the theory contains
five $\phi^4$ coupling constants as well as the standard $\phi^4$
coupling constant. They also pointed out that
we meet singular fluctuation on the critical surface
below dimension eight when we take
the zero-replica limit.  In order to resolve
this problem, they redefined the coupling constants such that
the beta functions for the new coupling constants do not
explicitly contain $n$, the number of the replica components.

The effective theory  contains not only the five $\phi^4$
coupling constants $u_i, \ (i=1, \ldots, 5)$, but also a parameter
$\Delta$ with the
mass dimension 2, which is related
to strength of the random field.  Since the new coupling constants
introduced in reference \cite{bd} are defined as linear combinations of
$\Delta u_i$,  they have the dimension $2+(4-d) = 6-d$ in
$d$-dimensional effective theory.

%Computing the scaling matrix (i.e., the derivative of  beta
%functions at a  fixed point) for the new coupling
%constants in the $6-\epsilon$ dimensions,   it was found that
%the scaling matrix has negative eigenvalues.
According to the beta functions corresponding to the new 
coupling constants,  there is a non-trivial fixed point in $6-\ep$ 
dimensions as in the case of the pure $\phi^4$ theory in $4-\ep$ 
dimensions.  However, aspects of flow is different from the pure theory 
because some flow is repelled by the fixed point.  
Therefore, they
concluded that the non-trivial fixed point
is unstable and the $\epsilon$-expansion cannot be carried out near  six
dimensions.  It indicates that the dimensional reduction \cite{aim,ps}
breaks down near the upper critical dimension as well as the
lower one \cite{i,bk}.

However,  it depends on choice of coupling constants in 
this model whether 
a fixed point is attractive or repulsive.  
Hence it is important to study how physical observables are expanded
by $u_i$ and $\Delta$, and to know what  the effective coupling
 constants suitable for perturbation are.    
To make this point clear,  it is instructive to consider 
the following simple example. Let $\lambda$ be a dimensionless, renormalized coupling constant.   Suppose that the beta function 
is given as 
\beq
	\be_\la(\la) = x \la - y \hat{\Delta} \la^2, 
\eeq
where $\hat{\Delta} \equiv \Delta m^{-2}$ and $m$ is the renormalized 
mass, the inverse of the correlation length.  We denote by $x$ and $y$ 
dimensionless coefficients. 
The fixed point $\la^*$ is 
easily found as $x/(y \hat{\Delta})$.   Then we have 
\beq
	\left. \frac{d \be_\la}{d \la} \right|_{\la=\la^*} = -x. 
\label{egn1}
\eeq
Next let us introduce a new coupling 
constant by 
\beq
	g \equiv \hat{\Delta} \la. 
\label{def_g}
\eeq
The beta function for $g$ is computed to be 
\beq
	\be_g(g) = m \frac{d g}{dm} = \frac{d g}{d \lambda} \beta_\la  - 2  g. 
\label{be_g}
\eeq
Using (\ref{def_g}) and (\ref{be_g}), one finds 
\beq
	\left. \frac{d \be_g}{d g} \right|_{g=g^*} = -x+2,  
\label{egn2}
\eeq
where $g^*$ is the zero of $\be_g$. 
Suppose that $0 < x < 2$.  Equation (\ref{egn1}) shows that 
the fixed point is unstable. On the contrary, it seems stable 
from the result (\ref{egn2}).  This example 
shows that we  have to clarify which coupling 
constants play a role of an expansion parameter for studying
the stability.  Note that the existence of the second term in 
the right-hand side of equation (\ref{be_g}) 
shows breakdown of the covariance of the beta function\cite{id}.  
It appears because the fixed dimensional parameter $\Delta$ 
enters the beta function and the definition (\ref{def_g}). 

  Although Br\'ezin and De Dominicis chose  $\Delta u_i$ 
for all $i$ as the coupling constants 
in order to eliminate $n$ dependence,  it has not yet
confirmed whether
the new coupling constants correspond to good expansion parameters.
The main purpose of this paper is to obtain 
the good expansion parameters and to reconsider the instability 
studied in reference \cite{bd}.  We conclude that their beta functions 
 are not for the good expansion parameters. 

This paper is organized as follows:
in the next section we perform the naive perturbative expansion of 
the free-energy density in the disorder phase 
and determine the expansion parameters.  Thanks to 
a finite mass in the disorder phase, there are no
singular behavior in the limit of $n \rightarrow 0$ discussed in 
 reference \cite{bd}. Therefore we can take the zero-replica limit 
 at the very beginning. 
We find that the good expansion parameters are obtained from 
the effective coupling constants 
$\Delta^{\al_i} u_i$, where $\al_i$ depends on $i$ 
(see equation (\ref{al})).  
Our effective coupling constants have various mass dimensions 
at the tree level.  One of them becomes marginal but the others 
are irrelevant in six dimensions.  When the dimensions lower 
by $\ep$, the marginal coupling constant becomes irrelevant 
at the non-trivial fixed point due to the one-loop correction, 
while the other coupling constants remain irrelevant.   

In section 3,  we turn to renormalized perturbation, where we 
define the renormalized effective coupling constants
and compute their beta functions.  Computing the 
scaling matrix (i.e., the derivative of beta functions at a 
fixed point),  we find that the fixed point is stable.

The correlation-length exponent 
and the anomalous dimension of the correlation function in $d=6-\ep$ are computed in section 4.  The results are identical  with the case of
the  $d=4-\ep$ pure $\phi^4$ theory
in the leading order of the $\ep$-expansion.
It is consistent with $(d, d-2)$ correspondence \cite{aim, ps}.

The last section is devoted to summary and future problems. 

\section{Coupling dependence of the free energy}
\label{coup}
Here we observe the bare-parameter
dependence of the free-energy density.

The $d$-dimensional
effective action derived by Br\'ezin and De Dominicis is given by
\beqa
 S = S_0 + S_{\rm int}, \nn\\
 S_0 \equiv \int d^d x \sum_{\al, \be =1}^n \frac{1}{2}\phi_{\al} (x)
 \left\{(- \partial^2 + t)\delta_{\al \be} -
  \Delta \right \} \phi_{\be} (x),  \nn\\
 S_{\rm int} \equiv \int d^dx \left(
 \frac{u_1}{4!} \sigma_{4} + \frac{u_2}{3!}
 \sigma_{3}  \sigma_{1}  +
    \frac{u_3}{8} \sigma_{2}^2 +
 \frac{u_4}{4}\sigma_{2}\sigma_{1}^2+
 \frac{u_5}{4!} \sigma_{1}^4  \right),
\label{s}
\eeqa
where $\al$ and $\be$, which run from 1 to $n$,
specify a replica component and
\beq
\sigma_{k}  \equiv  \sum_{\al=1}^{n} \( \phi_{\al}(x) \)^k,
\qquad (k=1, \ldots,  4).
\eeq
The interaction conjugate to $u_1$ contains one summation over
the replica index while the other interactions have more than one,
which apparently shows that the other coupling constants
$u_2, \ldots, u_5$
are less relevant in the zero-replica limit.

  The quadratic part $S_0$ defines
the propagator:
\beqa
 \hat G_{\al \be} (p)  &=& \frac{1}{p^2 + t} \de_{\al \be}
 + \frac{\Delta}{(p^2 + t)^2}
 + {\mbox{O}}(n), \nn\\
 &\equiv& A(p) \de_{\al \be} + B(p) + {\mbox{O}}(n).
\label{G}
\eeqa
Note that $B(p)$ dominates over $A(p)$ in low-momentum region with
small $t$, while $A(p)$ can be more relevant in the replica limit
because $\de_{\al \be}$ associated with it reduces  powers of $n$.

A simple dimensional analysis gives that the parameters
of this theory have the following mass dimensions
\beq
 [u_i] = 4-d \ (i = 1, \ldots, 5),  \ \ [t] = [\Delta]=2.
\eeq
We note that 
%physical observables should be measured
%by $t$ instead of $\Delta$
%because 
equation (\ref{G}) indicates that the correlation length $\xi$ is
proportional to $t^{-1/2}$ at the tree level.   We can see the 
critical phenomena by one-parameter tuning $t=0$ despite that 
$\Delta$ is a relevant coupling constant.    

Let us look at the perturbative expansion of the
free-energy density $\bar{f}$, which is obtained as
\beq
 \bar{f} = \lim_{n\rightarrow 0} \frac{f_{\rm rep}}{n},  \ \
 e^{-V f_{\rm rep}} = \int \prod_{\al =1}^n {\cal D} \phi e^{-S},
\eeq
where $V$ is the volume of the system.
The leading order is easily computed. The result is  written as
\beq
 \bar{f} =
 \frac{1}{2}\(1 - \Delta \frac{d}{d t} \)
 \int_{|p| \leq \La}
 \frac{d^d p}{(2\pi)^d} \log \(p^2+t\). 
\eeq
Here $\La$ is the ultraviolet cutoff. 
The integration is identical with the free-energy of
the Gaussian model, whose singular part behaves as $t^{d/2}$
when $t$ is sufficiently small. Hence the singular part of $\bar{f}$
is
\beq
 \bar{f}_{\rm sing} \propto \Delta t^{d/2-1}
 (1 + {\rm O}(\Delta^{-1} t)),
\eeq
where the
${\rm O}(\Delta^{-1} t)$ term is irrelevant near the
criticality:  $t \sim 0$ with fixed $\Delta$.

Next, we compute the first-order correction
\beq
 \lim_{n\rightarrow 0} \frac{1}{n V} \bra S_{\rm int} \ket. 
\eeq
For instance, the $u_1$ interaction gives
\beq
 \frac{u_1}{8} \({\cal A}_1 + \Delta{\cal A}_2 \)^2,
\label{u_1}
\eeq
where
\beq
 {\cal A}_k \equiv \int \frac{d^dp}{\(2\pi\)^d}
 \frac{1}{\(p^2 + t\)^k}, \ \ k=1,2.
\eeq
Extracting the singular part of 
${\cal A}_k$ from equation (\ref{u_1}), we
have
\beq
 \frac{u_1}{8} \(a_1t^{d/2-1} + \Delta a_2 t^{d/2-2} \)^2 =
 \frac{u_1}{8} \Delta^2 a_2^2 t^{d/2-3}
 \(1+ {\rm O}\(\Delta^{-1} t\) \), 
\eeq
where $a_1$ and $a_2$ are some constants. 
Similarly, collecting the leading contribution from
the each vertex that survives in the zero-replica limit,
we get
\beq
 \fl \bar{f}_{\rm sing} =
 \Delta t^{\frac{d}{2}-1}\left\{a_0
 + \( \frac{\Delta u_1}{8} a_2^2 t^{\frac{d}{2}-3} +
   \frac{2 u_2 +u_3}{2}  a_1 a_2t^{\frac{d}{2}-2}
 + \frac{\Delta^{-1}u_4}{2} a_1^2 t^{\frac{d}{2}-1}\) \right\},
\label{1st}
\eeq
up to the first order of $u_i$.  
Similar computation to the first few loops implies that 
the expansion parameters
are $\Delta^{\al_i} u_i t^{d/2-2-\al_i}$.  Here
%$\al_i$ denotes the power of $\Delta$ associated with $u_i$. 
%Explicitly, 
\beq
 \al_1 = 1, \qquad \al_2 =\al_3 = 0, \qquad \al_4 = -1, \qquad
 \al_5 = -2.
\label{al}
\eeq

We can see  that
the $\Delta u_1$ term survives while
the other corrections in (\ref{1st}) vanish in $6-\ep$ dimensions 
near the criticality.

%Note that the $u_1$ correction only remains in
%$d=6-\ep$ with small $\ep>0$. The other corrections vanish
%near the criticality.  It reflects that the right expansion parameters
%are not $\Delta u_i t^{d/2-3}$ for all $i$.

%As shown below, going to the higher order correction, we can see that
%the expansion parameters are (\ref{prms}) plus
%\beq
% \Delta^{-2} u_5 t^{-d/2}
%\label{u_5}
%\eeq
Do our expansion parameters work for all diagrams in 
$\bar{f}_{\rm sing}$? 
Let us  consider an arbitrary connected diagram in $\bar{f}$.
It contains two kinds of internal lines,  $A(p)$ and $B(p)$. Let
$\#A$ and $\#B$ be the number of $A(p)$ and $B(p)$ in the diagram 
respectively.
Since  $B(p)$ is proportional to  $\Delta$, the diagram
generates the following correction:
\beq
  F(t, \La) \Delta^{\# B} \prod_{i=1}^5 u_i^{p_i},  
%u_1^{p_1} u_2^{p_2} u_3^{p_3} u_4^{p_4}  u_5^{p_5} \Delta^{\# B} F(t).
\label{gen}
\eeq
where $F(t, \La)$ represents momentum integrations and
summations over the replica indices.
We find from the dimensional analysis that
 (\ref{gen}) has the following form:
%\beq
%	F(1, \La/t)  t^{d/2} \(\Delta t^{-1}\)^{\# B} 
%	\prod_{i=1}^5 \( u_i t^{d/2-2} \)^{p_i}.  
%% \fl t^{d/2} \( u_1 t^{d/2-2}\)^{p_1} \(u_2 t^{d/2-2}\)^{p_2}
%%  \(u_3 t^{d/2-2}\)^{p_3}\(u_4 t^{d/2-2}\)^{p_4}
%% \(u_5 t^{d/2-2}\)^{p_5} \(\Delta t^{-1}\)^{\# B},
%\label{gen2}
%\eeq

%Let $\al_i$ be the powers of $\Delta$ associated with  $u_i$
%in (\ref{prms}) and (\ref{u_5}).  That is,
%\beq
% \al_1 = 1, \qquad \al_2 =\al_3 = 0, \qquad \al_4 = -1, \qquad
% \al_5 = -2.
%\label{al}
%\eeq
%We can rewrite (\ref{gen2}) as
\beq
	F(1, \La/t)  \Delta t^{d/2-1} \( \Delta^{-1} t \)^{p_0}
	\prod_{i=1}^5 \( \Delta^{\al_i} u_i t^{d/2-2 - \al_i}\)^{p_i}. 
% \fl t^{d/2} \( \Delta u_1 t^{d/2-3}\)^{p_1} \(u_2 t^{d/2-2}\)^{p_2}
% \(u_3 t^{d/2-2}\)^{p_3} \(\Delta^{-1} u_4 t^{d/2-1}\)^{p_4}
%  \(\Delta^{-2} u_5 t^{d/2}\)^{p_5} \tilde{F}(\Delta^{-1} t). 
\label{gen3}
\eeq
Comparing (\ref{gen}) and (\ref{gen3}) we find 
\beq
 p_0 = \( \sum_{i=1}^5 \al_i  p_i \)  + 1 - \# B. 
\eeq
%
%$\tilde{F}(x)$ is the polynomial of $x$ with the degree of
%\beq
% \( \sum_{i=1}^5 \al_i  p_i \)  + 1 - \# B \geq 0.
%\eeq
We must show that 
\beq
	p_0 \geq 0. 
\label{p0}
\eeq
If  $p_0 < 0$, the expansion may fail because the factor 
$\Delta t^{-1}$ grows near the 
criticality, even though 
the last factor gives small correction.  

Here we prove (\ref{p0}).   Because the number of the internal
 lines are $2 \sum p_i$,
\beq
 \#A + \#B = 2\sum_{i=1}^5 p_i.
\label{total}
\eeq
From equation (\ref{s}) we find that 
the interaction conjugate to $u_i$ 
has $2-\al_i$ summations
of  the replica indices. Thus the number of the sums in $F$ is
\beq
 \sum_{i=1}^5 \( 2-\al_i \) p_i.
\label{repnum}
\eeq
It should be reduced up to 1 by Kronecker's delta
associated with $A(p)$: otherwise, $F$
becomes ${\mbox{O}}(n^2)$
and disappears in equation (\ref{1st}).
Thus, we have the following restriction to $\#A$:
\beq
 \# A \geq \sum_{i=1}^5 \( 2-\al_i \) p_i  -  1.
\label{A}
\eeq
Using (\ref{total}) and (\ref{A}), we obtain (\ref{p0}).  
%In other words, from equation (\ref{total}),
%\beq
% \# B \leq
% \( \sum_{i=1}^5 \al_i  p_i \)  + 1.
%\label{B}
%\eeq
%Thus the proof is completed.

\section{Renormalized perturbation}
Now we are going to the renormalized perturbation.
Using the renormalized 2-point vertex function
$\Gamma_{2 \al \be}(p)$,
we introduce the renormalized parameters $m$ and $\Delta_{\rm R}$ as
\begin{eqnarray}
 \Gamma_{2 \, \al \be} (p) &=&
 \( p^2 +m^2 \) \de_{\al \be} + {\Delta_{\rm R}} m^2 +
 {\rm O}(p^4).
 \label{rc1}
\end{eqnarray}
We have shown that the expansion parameters are
$\Delta^{\al_i} u_i t^{d/2 -2 - \al_i}$,  
%and $\Delta^{-1} t$,
so that we put the following
renormalization prescription by using the renormalization
counterparts for them:
\beqa
 \Delta_{\rm R}^{\al_i} \Gamma_{4 \, i} (0) m^{d-4} &=& g_i,
  \qquad (i=1, \ldots, 5), 
%\nn\\
% \Delta_{\rm R}^{-1} &=& g_0,
 \label{rc2}
\eeqa
where $\Gamma_{4 \, i} (p)$ is the renormalized 4-point functions
associated with $u_i$. Then one can perform the renormalized perturbation
with
the
small parameters $g_i$.

\begin{figure}[t]
  \begin{center}
	\setlength{\unitlength}{1mm}
	\begin{picture}(50,30)(0,0)
	\put(0,0){\includegraphics*[height=25mm]{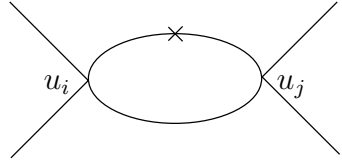}}
	\put(36,13){$u_j$}
	\put(5,13){$u_i$}
    \end{picture}
  \end{center}
\caption{Feynman diagram for the beta functions (\ref{beta_g}).}
\label{fig_beta}
\end{figure}

The one-loop
beta functions for $g_i$ in $6-\ep$ dimensions are easily computed
by the usual manner.   
%In order to make the difference
%from reference \cite{bd} clear, 
Following reference \cite{bd}, 
we keep the terms proportional to
\beq
 \int \frac{d^dp}{(2\pi)^d} \frac{\Delta_{\rm R}}{(p^2 + m^2)^3},
\eeq
which are expected to be important in $6-\ep$ dimensions. 
See figure \ref{fig_beta}. Here an internal line with a cross(x) 
represents 
$B(p)$ and without crosses $A(p)$. 

Similar calculation performed in reference \cite{bd} gives 
\beqa
 \be_1 &=& -\ep g_1 + 3
 K g_1^2,  \nn\\
 \be_2 &=& (2-\ep) g_2 + 3
 K g_1 (g_2 + g_3), \nn\\
 \be_3 &=& (2-\ep) g_3 +5
 K g_1 (g_2 + g_3), \nn\\
 \be_4 &=& (4-\ep) g_4 + 4
 K (g_2 + g_3)^2, \nn\\
 \be_5 &=& (6-\ep) g_5 + 36
 K (g_2+ g_3) g_4,
\label{beta_g}
\eeqa
where $K=\Gamma(4-d/2)/(4\pi)^{d/2}$.
The non-trivial fixed point is, up to ${\rm O}(\ep)$,  
\beq
 g_{1}^*  = \frac{1}{3K}\ep, \ \ g_{2}^* = \cdots g_{5}^* = 0.
\label{ntfp}
\eeq
Now we compute eigenvalues of the scaling matrix
\beq
 \left. \frac{\del \be_i}{\del g_j} \right|_{g=g^*}.
\eeq
The result is
\beq
 \{ \ep, \ 2-\ep, \ \frac{2}{3}(3+\ep), 4-\ep, \ 6-\ep \}.
\eeq
The eigenvalues are all positive. 
Thus the non-trivial fixed point is stable,
contrary to the conclusion in reference \cite{bd}. 

The discrepancy stems from the ways of redefinition of the 
coupling constants. Br\'ezin and de De Dominicis have essentially 
chose $\Delta u_i$ as the effective coupling constants.  
The linear terms of the beta functions correspond to $\Delta u_i$ 
are $-\ep$ for all $i$, and negative eigenvalues emerge. 
However, 
the effective coupling constants that appear in perturbative 
expansion are $\Delta^{\al_i} u_i$ as we have seen in the 
previous section. Since $\Delta$ has the finite mass dimension, 
scaling behavior of $\Delta u_i$ does not reflect the stability 
of the fixed point.

\section{The $\epsilon$-expansion near six dimensions}

Finally, we compute the leading order of the $\epsilon$-expansion 
for the
critical exponents $\nu$ and $\eta$
at the non-trivial fixed point (\ref{ntfp}).

\begin{figure}[b]
  \begin{center}
	\setlength{\unitlength}{1mm}
	\begin{picture}(80,20)(0,0)
	\put(0,0){\includegraphics*[height=20mm]{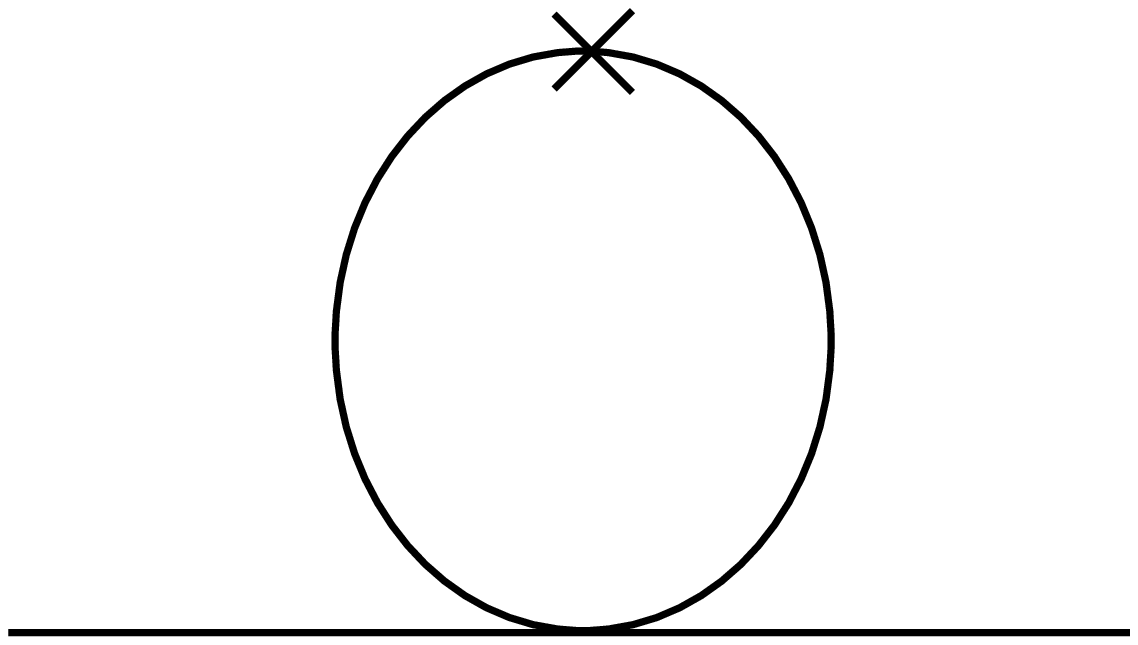}}
	\put(45,-9){\includegraphics*[height=25mm]{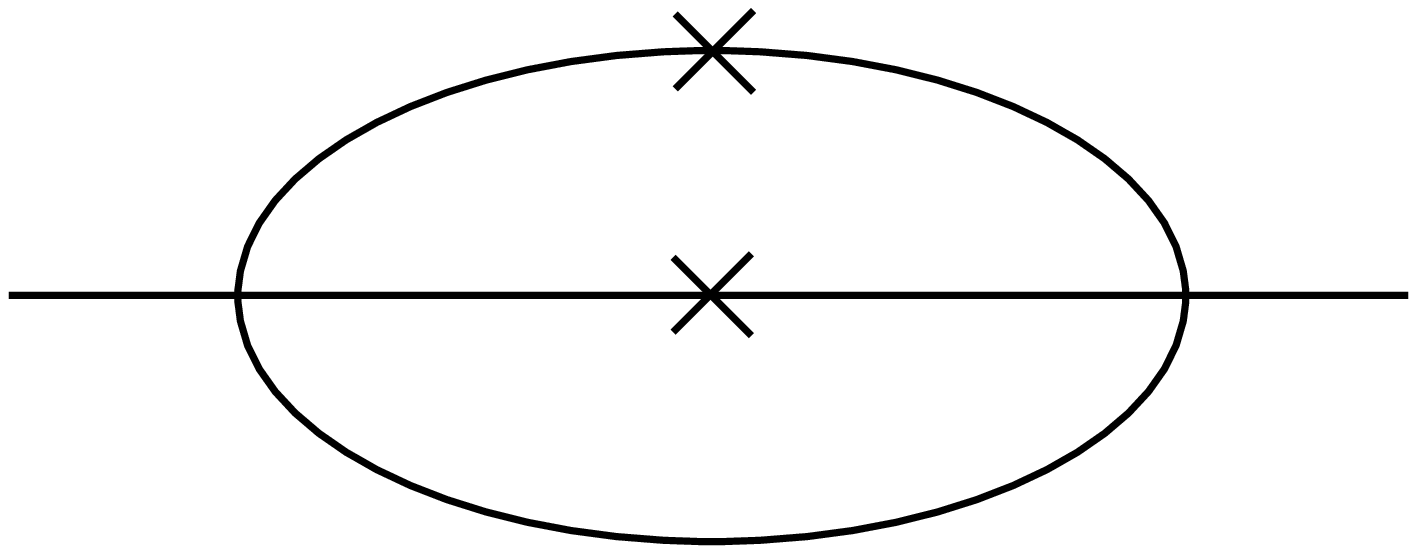}}
	\put(79,5){$u_1$}
	\put(48,5){$u_1$}
	\put(16.5,-3){$u_1$}
    \end{picture}
  \end{center}
\caption{Feynman diagrams contributing to the self-energy.}
\label{self-energy}
\end{figure}

To compute the critical exponents one can take account of
the leading infrared divergence only.
Figure \ref{self-energy} represents
the most singular Feynman diagrams
contributing to the self-energy. 
One easily sees that only the $u_1$ term
contributes to the most singular diagrams;
vertices $u_2,\ldots ,u_5$ combined with $B(p)$
giving additional $n$ factor are suppressed by the
$n \rightarrow 0$ limit.
%\begin{figure}
%\caption{Feynman diagrams contributing to the self-energy.}
%\label{self-energy}
%\end{figure}
Thus, from the renormalization conditions (\ref{rc1}) and (\ref{rc2}),
the one-loop correction of the bare mass and
the two-loop computation of the field renormalization factor $Z_1$
are written in terms of the renormalized parameters as follows:
\begin{eqnarray}
t = m^2 \biggl\{
1 - \frac{1}{2} \frac{\Gamma (2-d/2)}{(4 \pi)^{d/2}} g_1
+ {\rm O} (g_1^2)
%- \frac{1}{4} \frac{\Gamma (3-d/2)}{(4 \pi)^{d/2}}
%\frac{\Gamma (2-d/2)}{(4\pi)^{d/2}} g_1^2
%+ \frac{1}{6} \Sigma (0,1) g_1^2
\biggr\},
\label{t}
\\
Z_1 =
\biggl\{
1 + \frac{1}{12} \frac{\Gamma(6-d)}{(4\pi)^d} g_1^2
\biggr\}^{-1}.
\end{eqnarray}
From equations (\ref{beta_g}) and (\ref{t}),   
the composite field renormalization factor $Z_2$
is written at the one-loop level as follows:
\begin{equation}
Z_2 = 
\left. \frac{d m^2}{d t} \right|_{\Delta,u_i \rm{: fixed}} = 
1 - \frac{1}{2} \frac{\Gamma(3-d/2)}{(4\pi)^{d/2}} g_1 + {\rm O} (g_1^2).
\end{equation}
Following the standard argument \cite{bgz,id},
define 
\begin{eqnarray}
\gamma_1(g_1,\ldots,g_5)
&\equiv&
m \frac{\partial}{\partial m} \ln Z_1
\biggr|_{\Delta,u_i \rm{: fixed}},\nonumber\\
\gamma_2(g_1,\ldots,g_5)
&\equiv&
m \frac{\partial}{\partial m} \ln Z_2
\biggr|_{\Delta,u_i \rm{: fixed}}. 
\end{eqnarray}
We can compute the critical exponents $\nu$ and $\eta$ 
from the following relationship:
%the dimensionless quantities
%$\gamma_1(g_1,\ldots,g_5)$ and $\gamma_2(g_1,\ldots,g_5)$
%are related to the critical exponents $\nu$ and $\eta$ through
\begin{eqnarray}
\nu &=&
\frac{1}{2 - \gamma_2(g_1^*,\ldots,g_5^*)},\nonumber\\
\eta &=&
\gamma_1(g_1^*,\ldots,g_5^*).
\end{eqnarray}
Substituting the values of the fixed point (\ref{ntfp}) into the above
equations, we obtain the leading order of the $\epsilon$-expansion for the
critical
exponents $\nu$ and $\eta$, 
\begin{eqnarray}
\nu &=&
\frac{1}{2} + \frac{1}{12} \epsilon + \rm{O}(\epsilon^2),\nonumber\\
\eta &=&
\frac{1}{54} \epsilon^2 + \rm{O}(\epsilon^3).
\end{eqnarray}
These are exactly the same as the leading-order results in the $d=4-\ep$
pure
$\phi^4$ theory \cite{aim,ps}.

\section{Summary and discussion}
In this paper, we have clarified effective coupling constants 
for perturbative expansion of physical observables in 
effective theory for the random field Ising model and 
reconsidered the stability discussed in reference \cite{bd}. 
Our result ensures the $\ep$ expansion near the upper critical 
dimension.  We have also calculated the exponents $\nu$ and $\eta$.   
The result is consistent with
the Parisi-Sourlas dimensional reduction in the leading-order computation.
It is a nontrivial problem whether the consistency is
preserved beyond the
leading order, which will be reported elsewhere. 

Our beta functions (\ref{beta_g}) tell us 
 that the upper critical dimension 
is six, which is consistent with the rigorous result \cite{t}.  
Further, they 
show that  $g_2$ and $g_3$ change  
to relevant coupling constants when $d$ becomes 
lower than four.   Thus it does not surprise us  that 
the dimensional reduction does not hold when $d=3$ \cite{bk}. \\[3mm]

\noindent
We are grateful to E. Br\'ezin for explaining their work and for his
criticism.  
We also 
would like to thank C. Itoi for valuable discussions and comments and 
for careful reading of the manuscript  .


\section*{References}
\begin{thebibliography}{<num>}
\bibitem{bd}
Br\'{e}zin E and De Dominicis C, {\it Europhys. Lett.} {\bf{44}} (1998) 13.
\bibitem{bgz}
Br\'ezin E, Le Guillou J -C and Zinn-Justin J,
{\it Phase Transitions and  Critical Phenomena vol.6},
C. Domb and M. S. Green eds. (Academic, New York 1976);
Parisi G, {\it Statistical Field Theory}, (Addison-Wesley, 1988). 
\bibitem{id}
Itzykson C and Drouffe J -M, {\it Statistical Field Theory},
(Cambridge University Press, 1989).
\bibitem{aim}
Aharony A, Imry Y and Ma S K, {\it Phys. Rev. Lett.} {\bf{37}} (1976) 1364.
\bibitem{ps}
Parisi G and Sourlas N, {\it Phys. Rev. Lett.} {\bf{43}} (1979) 744;
{\it Nucl. Phys.} {\bf{B206}} (1982) 321.
\bibitem{i}
Imbrie J Z, {\it Phys. Rev. Lett.} {\bf{53}} (1984) 1747;
{\it Commun. Math. Phys.} {\bf{98}} (1985) 145.
\bibitem{bk}
Bricmont J and Kupiainen A, {\it Phys. Rev. Lett.} {\bf{59}} (1987) 1829;
Commun. Math. Phys. {\bf{116}} 539;
see also,
Aizenman M and Wehr J, {\it Phys. Rev. Lett.} {\bf{62}} (1989) 2503;
{\it Commun. Math. Phys.} {\bf{130}} (1990) 489;
Wehr J and Aizenman M, {\it J. Stat. Phys.} {\bf{60}} (1990) 287.
%\bibitem{wk}
%Wilson K G and Kogut J, {\it Phys. Rep.} {\bf{12C}} (1974) 75.
\bibitem{t}
Tasaki H, {\it J. Stat. Phys.} {\bf{54}} (1989) 163.
\end{thebibliography}
\end{document}